\documentclass[a4paper,10pt]{article}\textheight 24.5cm \textwidth 17cm\voffset=-1.in\hoffset= - 1.in

\def\tr{{\textrm{tr}}}

\usepackage[dvips]{graphicx}
\usepackage{psfrag}
\usepackage{amsmath}
\usepackage{graphicx}
\usepackage{amsfonts}
\usepackage{amssymb}
\usepackage{mathtext}         %
\usepackage[T2A]{fontenc}     %
\usepackage[cp1251]{inputenc} %

\def\cP{{\mathcal{P}}}

\def\tr{{\textrm{tr}}}

\begin{document}


\vspace{0.5cm}

\begin{center}
\begin{LARGE}
{\Large \textbf{Two dimensional gravity in genus one in Matrix
Models,}}

{\Large\textbf{Topological and Liouville approaches.}}

\vspace{0.3cm}

\end{LARGE}

\vspace{0.5cm}

\begin{large}

\textbf{A.Belavin}, $\;$\textbf{G.Tarnopolsky}

\end{large}
\large

 \vspace{1.cm}

L.D. Landau Institute for Theoretical Physics\\
  Chernogolovka, 142432, Russia\\

\vspace{1.0cm}
\begin{center}
   \textbf{Abstract}
\end{center}

\begin{flushleft}
    One-matrix model in $p$-critical point on torus is considered.
The generating  function of correlation numbers in genus one is evaluated
and used for computation correlation numbers in KdV and CFT frames.
 It is shown that the correlation numbers in KdV frame in genus one
 satisfy the Witten topological gravity recurrence relations.
\end{flushleft}

\vspace{.8cm}

\parbox{11cm}

\end{center}

\large

\section{Introduction}

There exist several different approaches to the 2d Quantum gravity.
One of them is the continuous approach. In this approach the theory
is determined by the functional integral over all metrics
\cite{polyakov}. Calculation of this integral in the conformal gauge
leads to the Liouville field theory. Therefore this approach is
called the Liouville gravity.

The other way to describe the sum over 2d surfaces is  the discrete
approach. It is based on the idea of approximation of
two-dimensional geometry by an ensemble of planar graphs of big
size. Technically the ensemble of graphs is usually defined by
expansion into a series of perturbation theory of the integral over
$N\times N$  matrixes. That is why this approach is called the
Matrix Models (further MM). References to the both approaches can be
found in the review \cite{review}.

After \cite{KPZ,K,review}  the coincidence of the gravitational
dimensions was found the conjecture that both of these approaches
describe the same variant of the 2d Quantum gravity appeared.
Therefore it was naturally to expect that the correlation numbers
will are also the same. However  the attempt of a naive
identification of the correlation numbers  doesn't lead to the
agreement in a general case.

In \cite{MSS, BZ2} a conjecture was proposed and checked  that there
exists a ``resonance'' transformation of coupling constants in
Matrix Models, from the standard definition (the so-called KdV frame) to
another (the so-called Liouville or CFT frame), such that new defined
correlation numbers of MM coincide  with naturally defined ones in
the Liouville Gravity.

The form of the transformation was conjectured in \cite{BZ2} for the
particular case of the $p$-critical  One-Matrix Model (OMM), which
corresponds to the Minimal Liouville gravity
$\mathcal{MG}_{2/2p+1}$. The conjectured identity of the correlation
numbers was checked up to five-point case in genus zero
\cite{BZ2,T}.

At last  the third approach~-- 2d Topological gravity was invented
by Witten in \cite{EW}, who built axiomatics of this theory along
the lines of intersection theory .  It was conjectured and checked
(for genus zero) in \cite{EW} that correlation numbers in
Topological gravity and in Matrix models coincide. It should be
mentioned that this fact takes place if correlation numbers in OMM
are calculated in KdV frame.

The article is organized in the following way. At first we review
the method of orthogonal polynomials for the solution of Matrix
model, the double scaling limit and Douglas sting equation. Then we
use these tools to compute the torus partition function in
$p$-critical One-matrix model .

We use the explicit expression for the partition function in genus
one to compute the correlation numbers in KdV, as well as in CFT
frames.

The results in CFT frame should be compared with the correlation
numbers in the Minimal Liouville gravity, which have not been
computed yet. In genus one we expect a coincidence similar to
that observed in genus zero \cite{MSS, BZ2}.

After that, we evaluate explicitely  the first three correlation
numbers in KdV frame. Afterwards we derive the recursion relation in
genus one and observe that it coincides with the one of Topological
gravity. Then we compute two first correlation numbers in Conformal
frame. At the end we discuss some open problems.

\section{The method of orthogonal polynomials}
Here we give some well-known aspects of the method of orthogonal
polynomials \cite{review}. The partition function which includes
surfaces of all genera is
\begin{align}
Z(v_{k},N ) = \log \int dM e^{-\tr V(M)}, \label{Z0}
\end{align}
where $M$ is hermitian $N\times N$ matrix and $V(M)
=N\sum_{k=1}^{p+1}v_{k}M^{2k} $ is the polynomial potential with
different coupling constants $v_{k}$, and in further calculation we
fix $v_{p+1}=\frac{(p+1)!p!}{(2p+2)!}$ for simplicity. On the other
hand, it is known \cite{review} that  partition function (\ref{Z0})
is expressed through the sum
\begin{align}
Z = \sum_{h=0}^{\infty} N^{2-2h}Z_{h}, \label{FZ}
\end{align}
where $h$ is genus of surface and $Z_{h}$ is the partition function
of all surfaces with genus $h$. In this paper we evaluate torus
partition function $Z_{1}$, expanding the integral (\ref{Z0}) in
$1/N$ series.

In this section we carry on evaluation using the method of
orthogonal polynomials.

Since the integrand in (\ref{Z0}) depends only on the eigenvalues of
the matrix $M$, we can factorize the integration measure into the
product of the Haar measure for unitary matrices and an integration
measure for eigenvalues.

Thus we have
\begin{align}
Z(v_{k},N ) = \log\int
\prod_{i=1}^{N}d\lambda_{i}\Delta^{2}(\lambda)e^{-\sum_{i}V(\lambda_{i})},
\label{Z1}
\end{align}
where $\lambda_{i}$'s are the $N$ eigenvalues of the hermitian
matrix $M$ and $ \Delta(\lambda) =
\prod_{i<j}(\lambda_{i}-\lambda_{j}) $ is the Vandermonde
determinant. Then we define the  set of orthogonal polynomials
$P_{n}(\lambda) = \lambda^{n}+...$, by
\begin{align}
\int_{-\infty}^{\infty} d\lambda
e^{-V(\lambda)}P_{n}(\lambda)P_{m}(\lambda) = s_{n}\delta_{nm}.
\end{align}
It is easy to see that the Vandermonde determinant can be written as
$ \Delta(\lambda)= \det P_{j-1}(\lambda_{i})$. After some
calculations we get the expression for general partition function
\begin{align}
&Z =N \sum_{k=1}^{N-1}(1-k/N)\log (s_{k}/s_{k-1}).
\end{align}
Therefore we need to evaluate the $s_{k}$.  Due to the assumption
that the potential $V(\lambda)$ is even, and also due to
orthogonality and normalizing condition of polynomials one can
obtain the simple recursion relation
\begin{align}
\lambda P_{k} = P_{k+1}+R_{k}P_{k-1} \label{rr},
\end{align}
where $R_{k}$ is constant. Then we can derive $ \int
e^{-V}P_{k}\lambda P_{k-1} d\lambda = R_{k}s_{k-1}=s_{k}$, and thus
$s_{k}/s_{k-1}=R_{k}$. After that the partition function writes as
follows
\begin{align}
&Z =N \sum_{k=1}^{N-1}(1-k/N)\log R_{k}. \label{E1}
\end{align}
The next step is the key relation that will allow us to determine
$R_{k}$ :
\begin{align}
ks_{k-1}=\int e^{-V}P'_{k}P_{k-1}=\int e^{-V}V'P_{k}P_{k-1}.
\label{keyr}
\end{align}
\noindent Since the derivative of the polynomial potential $V(\lambda)$ is
equal $V'(\lambda) =  \sum_{k=1}^{p+1}2kv_{k}\lambda^{2k-1}$,
we need to evaluate integrals like $ \int e^{-V}\lambda^{2n-1}P_{k}P_{k-1}$.

To do this one should apply relation (\ref{rr}) precisely $2n-1$ times:
\begin{multline}
\lambda^{2n-1}P_{k}=\lambda^{2n-2}(P_{k+1}+R_{k}P_{k-1})=\\=
 \lambda^{2n-3}(P_{k+2}+R_{k+1}P_{k}+R_{k}P_{k}+R_{k}R_{k-1}P_{k-2})=...=P_{k+2n-1}+....
 \notag
\end{multline}
But from this sum only the terms with $P_{k-1}$   will give
contribution to integral (\ref{keyr}).  The contributions of the
such terms to the integral (\ref{keyr}) may be visualize paths  \,of
$2n-1$ steps ($n-1$ steps up and $n$ steps down) starting at $k$ and
ending at $k-1$.

Each step down from $m$ to $m-1$ receives a factor of $R_{m}$ and
each step up receives a factor of unity.

The total number of the paths is given by the binomial
coefficient $C_{2n-1}^{n}$  and each path  gives a contribution to the factor of $s_{k-1}$
arising  due to
the integral $\int e^{-V}P_{k-1}P_{k-1}$.
Thus for our potential $V(\lambda)$ one can obtain the equation
\begin{align}
\frac{k}{N}=\tilde{W}(R_{k},R_{k\pm 1},...,R_{k\pm p}), \label{E2}
\end{align}
where the function $\tilde{W}$ is the polynomial of $R$-terms. In
explicit form it can be written like
\begin{align}
\tilde{W}(R_{k},R_{k\pm 1},...,R_{k\pm p}) = \sum_{n=0}^{p+1}2nv_{n}\tilde{W}_{n}
=\sum_{n=1}^{p+1}2nv_{n}\sum_{\{\sigma_{2n-1}\}}
R_{k+m_{1}}\cdot...\cdot R_{k+m_{n}},\label{TW}
\end{align}
where $\{\sigma_{2n-1}\}$ denotes all paths of length $2n-1$ of the
type, which is described above. Each path determines the numbers
$m_{1},...,m_{n}$, where $k+m_{i}$ are the coordinates of points,
from which the path goes down. Solving (\ref{E2}) with respect to
the $R_{k}'$s, and putting the answer in (\ref{E1}) we obtain full
partition function $Z$. In the next part we will carry on this
procedure using perturbation theory with parameter $1/N$.

\section{Evaluation of $Z_{0}$ and $Z_{1}$}
We are going  to find $R_{k}(N)$ which satisfy to equation
(\ref{TW}) when $N$ goes to infinity. Therefore we can look for this
solution in terms of the smooth function   $R(\xi,N)$ of variable
$\xi \in [0,1]$,  such that $R(\frac{k}{N},N)=R_{k}$.

It means that $R(\xi+m/N,N)$ has the Taylor expansion
\begin{align}
R(\xi+m/N,N) =
R(\xi,N)+\frac{m}{N}R_{\xi}(\xi,N)+\frac{m^{2}}{2N^{2}}R_{\xi\xi}(\xi,N)+O\big(\frac{1}{N^{3}}\big),
\end{align}
where $R_{\xi} \equiv \frac{\partial R}{\partial \xi}$ and $R_{\xi
\xi} \equiv \frac{\partial^2 R}{\partial \xi^2}$.

In addition we assume that  the function $R(\xi,N)$ itself can be also expanded in series
\begin{align}
R(\xi,N) =
R(\xi)+\frac{1}{N}R_{1}(\xi)+\frac{1}{N^{2}}R_{2}(\xi)+...,\label{RE}
\end{align}
when $N$ goes to infinity.

Taking into account the first assumption we get from (\ref{TW})
after some calculations (see appendix B) the following expansion of
$\tilde{W}$:
\begin{align}
\tilde{W}(R_{k},R_{k\pm 1},...) = W(R(\xi,N)) +
\frac{1}{N}W_{1}(R(\xi,N))+\frac{1}{N^{2}}W_{2}(R(\xi,N))+O\big(\frac{1}{N^{3}}\big),
\end{align}
where
\begin{align}
&W(R(\xi,N))
=\sum_{n=1}^{p+1}\frac{(2n)!}{n!(n-1)!}v_{n}R^{n}(\xi,N),\\
&W_{1}(R(\xi,N))=0,\\
&W_{2}(R(\xi,N))=\frac{RR_{\xi\xi}}{6}W''(R(\xi,N))
+\frac{RR_{\xi}^{2}}{12}W'''(R(\xi,N)),
\end{align}
and  $W'(R)\equiv \frac{dW}{dR}$ , $W'''(R)\equiv \frac{d^3W}{dR^3}$. \\

Thus the equation (\ref{E2}) takes a form
\begin{align}
\xi=W(R(\xi,N))+\frac{RR_{\xi\xi}}{6N^{2}}W''(R(\xi,N))
+\frac{RR_{\xi}^{2}}{12N^{2}}W'''(R(\xi,N))+O\big(\frac{1}{N^{4}}\big),
\label{W00}
\end{align}
where $\xi = k/N$.

Using the expansion (\ref{RE}) we have from
(\ref{W00}) at $N \to \infty$
\begin{align}
&\xi =W(R(\xi)), \label{eqxi}\\
&R_{1}(\xi)=0, \notag \\
&R_{2}(\xi)=
-\frac{R(\xi)}{12W'(R(\xi))}\left(2R_{\xi\xi}W''(R(\xi))
+R_{\xi}^{2}W'''(R(\xi))\right). \label{Ans}
\end{align}

Let us come back to  partition function (\ref{E1}). In order to
obtain the coefficients $Z_{h}$ from (\ref{FZ}) we need to expand
the sum in formula (\ref{E1}) in $1/N$ series. We pass from discrete
sum of number $k$ to integral of continious variable $\xi$. For this
procedure  one need to use Euler-Maclorein formula (see Appendix A).
\begin{align}
Z =N^{2} \int_{0}^{1} d\xi(1-\xi)\log R(\xi,N) -
\frac{N}{2}(F(1)-F(1/N))+\frac{1}{12}(F'(1)-F'(1/N))+O(1/N),
\label{ZEX}
\end{align}
where $F(\xi)=(1-\xi)\log R(\xi,N)$.  Using the formulae (\ref{RE})
and (\ref{Ans}) we obtain from (\ref{ZEX})
\begin{align}
&Z_{0} =  \int_{0}^{1} d\xi(1-\xi)\log R, \label{Z0I} \\
&Z_{1} =-\frac{1}{12} \int_{0}^{1}d\xi
(1-\xi)\frac{2R_{\xi\xi}W''(R) +R_{\xi}^{2}W'''(R)}{W'(R)},
\label{Z1I}
\end{align}
where $R=R(\xi)$ is the solution of equation (\ref{eqxi}).

In (\ref{Z0I}, \ref{Z1I}) we keep only first integral term in r.h.s.
of (\ref{ZEX}) and omit the others. The reason is that in the
vicinity of critical point, which we will be interesting below,
these terms have less singularity than the integral.

\section{The vicinity of $p$-critical point}
The $p$-critical
point are defined by the system of equations
\begin{align}
W(R_{c})=1, \quad W'(R_{c})=0,\quad ... \quad W^{(p)}(R_{c})=0.
\label{at1}
\end{align}
Actually it is the system of equations, which determine coefficients
$v_{k}^{c},\; k=1,...,p$, and define the $R_{c}$. Thus if we put the
$v_{k}=v_{k}^{c}$, then we have from (\ref{eqxi})
\begin{align}
\xi = 1 +(R-R_{c})^{p+1}.
\end{align}
Then if we consider the special small deviations $\delta
v_{k}=v_{k}-v_{k}^{c}$, such that  equations in (\ref{at1}) take a
form
\begin{align}
 W(R_{c})=1+t_{p-1}, \quad W'(R_{c})=t_{p-2},\quad ...\quad
W^{(p-1)}(R_{c})=t_{0}, \quad W^{(p)}(R_{c})=0.
\end{align}
Denoting $u=R-R_{c}$ one can obtain from (\ref{eqxi})
\begin{align}
\xi =W(u)= u^{p+1}+t_{0}u^{p-1}+\sum_{k=1}^{p-1}t_{k}u^{p-k-1}+1.
\label{Wx}
\end{align}
We can rewrite (\ref{Wx}), making a substitution $\xi = 1-y$, as
follows
\begin{align}
&\cP(u)+y=0, \label{FormPu}
\end{align}
where
\begin{align}
\cP(u) \stackrel{\textrm{def}}{=}
u^{p+1}+t_{0}u^{p-1}+\sum_{k=1}^{p-1}t_{k}u^{p-k-1} \label{P}.
\end{align}
In view of formula (\ref{FormPu}), in further we treat the variable
$u$, as the function of variables $\{t_{k}\}$ and $y$:\,
$u=u(\{t_{k}\},y)$ (or $u(y)$ for short).

Therefore from (\ref{Z0I})
and (\ref{Z1I})  one can get for singular part of  the partition function \cite{review} of genus zero and genus one
\begin{align}
&Z_{0}=\frac{1}{R_{c}}\int_{0}^{1}dy\,y\, u(y), \label{Z00}\\
&Z_{1}= -\frac{1}{12}\int_{0}^{1}dy\,y\left(\frac{2\cP''(u)u_{yy}
+\cP'''(u)u_{y}^{2}}{\cP'(u)}\right) \label{Z1M},
\end{align}
where $\cP'(u) \equiv \frac{d\cP}{du}$, $\cP''(u) \equiv
\frac{d^{2}\cP}{du^{2}}$, $\cP'''(u) \equiv
\frac{d^{3}\cP}{du^{3}}$.

The singularity of $Z_{1}$ arises due to the contribution
of the neighbourhood  of the point $y=0$ in the integral.

Using  (\ref{P}) in relations $u_{y} = -1/\cP'$ and
$u_{yy}=-\cP''/(\cP')^{3}$,  we obtain the following final answers
for the singular  parts of the partition   functions $Z_{0}$ and
$Z_{1}$
\begin{align}
&Z_{0}=\frac{1}{2}\int_{0}^{u^{*}}\cP^{2}(u)du,\label{Z0F}\\
&Z_{1}= -\frac{\log \cP'(u^{*})}{12}, \label{Z1F}
\end{align}
where $u^{*}=u^{*}(t_{0},t_{1},...,t_{p-1})$ is the suitably chosen
root \cite{BZ2} of the polynomial $\cP(u)$.

In further we want to evaluate correlation numbers in the critical
point $(t_{0},0,...,0)$. Let us denote the cosmological constant
$t_{0}=\mu$. The deviation from critical point is defined by
parameters $t_{k},\;k=1,..,p-1$.  In the vicinity of the  critical
point $\mu \ll 1$, $t_{k}\sim \mu^{\frac{k+2}{2}}$, the scaling
partition function, which corresponds to the surfaces of genus $h$,
is scale invariant
\begin{align}
Z_{h}[\lambda\mu,\lambda^{\frac{k+2}{2}}t_{k}]
=(\lambda^{p+3/2})^{1-h}Z_{h}[\mu,t_{k}]. \label{scale}
\end{align}

Double scaling limit corresponds to $N \to \infty$, while $\mu$ and
$t_{k} \to 0$ proportionally
$(N^{2}\varepsilon^{2})^{-\frac{2}{2p+3}}$ and
$(N^{2}\varepsilon^{2})^{-\frac{k+2}{2p+3}}$ correspondingly, where
$\varepsilon$ is some finite parameter.

Making suitable replacement of variables, using the rescaling (\ref{scale}) and performing the
substitution $Z/N^{2}\varepsilon^2 \to Z$ for simplicity, we arrive to the
expression for the partition function in the double scaling limit
$Z[\mu,t_{k},\varepsilon]$
\begin{align}
Z[\mu,t_{k},\varepsilon] =  \sum_{h=0}^{\infty}
\varepsilon^{2h}Z_{h}[\mu,t_{k}],
 \label{Ze}
\end{align}
where $\varepsilon$  is the parameter, which is responsible for
genus expansion. The expansion (\ref{Ze}) is similar to the ones in
Liouvile gravity.

In the next section we derive the same expression for  the torus
partition function by method different to ones in section 2. Namely
we will use the string equation  for the partition function
$Z[\mu,t_{k},\varepsilon]$ and the expansion respect to the small
parameter $\varepsilon$.

\section{String equation}
In this section we  show how to obtain the expressions (\ref{Z0F})
and (\ref{Z1F}) using the double scaling limit and the Douglas
string equation. The string equation is the equation for function
$u(x,\varepsilon,\mu,t_{k})$ (or $u(x,\varepsilon)$ for short),
which is connected with  the partition function in the double
scaling limit $Z[\mu,t_{k},\varepsilon]$ as
\begin{align}
u(x,\varepsilon)=\frac{d^{2}Z}{dx^{2}}, \label{ZSE}
\end{align}
and looks as follows
\begin{align}
[\hat{P},\hat{Q}]=1, \label{SQ}
\end{align}
where $\hat{Q}=\varepsilon^{2}d^{2}+u(x)$ and $\hat{P}
=-\sum_{k=1}^{p+1}t_{p-1-k}\hat{Q}^{k-1/2}_{+}$ are two differential
operators.

\noindent $\hat{Q}^{k-1/2}_{+}$ is the non-negative part of the
pseudo-differential operator $\hat{Q}^{k-1/2}$.

In view of (\ref{Ze}) we  look for $u(x)$ in the form
\begin{align}
u(x,\varepsilon) = \sum_{h=0}^{\infty}\varepsilon^{2h}u_{h}(x),
\label{uex}
\end{align}
where, obviously, $u_{h}$
\begin{align}
u_{h}(x)=\frac{d^{2}Z_{h}}{dx^{2}}. \label{ZSEh}
\end{align}
It is known \cite{review}, that
\begin{align}
 [\hat{Q}^{k-1/2}_{+},\hat{Q}]=\frac{dS_{k}}{dx},
\end{align}
where the coefficients $S_{k}(u)$ obey the recursion relation
\begin{align}
\frac{dS_{k+1}}{dx}= u\frac{dS_{k}}{dx}+\frac{1}{2}u_{x}S_{k}
+\frac{\varepsilon^{2}}{4}\frac{d^{3}S_{k}}{dx^{3}}, \label{RR}
\end{align}
with the boundary conditions $S_{0}=\frac{1}{2}$ and $S_{k}$($k\neq
0$) vanish at $u=0$, and we assume that $ u_{x}=
\partial u/\partial x, \; u_{xx} = \partial^{2}u/\partial x^{2} $.
Thus we obtain from (\ref{SQ}) the relation
\begin{align}
\sum_{k=1}^{p+1}t_{p-1-k}S_{k}(u) =-x.  \label{SQF}
\end{align}

The solution of  the recursion relations (\ref{RR}) (see Appendix С)
, including the first three terms is
\begin{align}
S_{k}(u)
=\frac{C_{2k}^{k}}{2^{2k+1}}\left(u^{k}+\frac{\varepsilon^{2}k(k-1)}{6}u^{k-2}u_{xx}
+\frac{\varepsilon^{2}k(k-1)(k-2)}{12}
u^{k-3}u_{x}^{2}\right)+O(\varepsilon^{4}). \label{Rk}
\end{align}
Thus after rescaling the parameter $t_{k} \to
\frac{2^{2k+1}}{C_{2k}^{k}}t_{k}$, we can obtain from (\ref{SQF})
that
\begin{align}
\cP(u)+\varepsilon^{2}\left(\frac{1}{6}\cP''(u)u_{xx}
+\frac{1}{12}\cP'''(u)u_{x}^{2}\right)=O(\varepsilon^{4}),
\label{DSF}
\end{align}
where $\cP(u)$ is the polynomial from (\ref{P}) and $x =t_{p-1}$,
$t_{-2}=1$,  $t_{-1}=0$.

Using the expansion (\ref{uex}), we get from (\ref{DSF})to the
zeroth order in the $\varepsilon$, that $u_{0}(x)$ obeys
\begin{align}
&\cP(u_{0})=0,
\end{align}
therefore
\begin{align}
u_{0}=u^{*}(t_{1},...,t_{p-2},x), \label{U0}
\end{align}
where $u^{*}$ is the suitably chosen root of the polynomial
$\cP(u)$. To the second order  in the  $\varepsilon$ gives for the
$u_{1}(t_{1},...,t_{p-2},x)$ the following expression
\begin{align}
&u_{1}=-
\frac{\cP'''(u^{*})(u^{*}_{x})^{2}+2\cP''(u^{*})u_{xx}^{*}}{12\cP'(u^{*})}.
\label{U1}
\end{align}
Knowing $u_{0}$ and $u_{1}$ we can find corresponding  the partition
functions $Z_{0}$ and $Z_{1}$, using (\ref{ZSEh}) and  the fact that
if $Z$ and $u^{*}$ are connected by relation
\begin{align}
\frac{\partial^{2} Z}{\partial x^{2}} = f(u^{*}), \label{UR1}
\end{align}
then
\begin{align}
Z = -\int_{0}^{u^{*}}\cP(u)\cP'(u)f(u)du. \label{ans1}
\end{align}
This formula can be checked by straightforward calculation.

Integrating by parts and omitting the regular terms, we get from
(\ref{ZSEh}), (\ref{U0}), (\ref{U1}) and (\ref{ans1})

\begin{align}
Z_{0}&=
 \frac{1}{2}\int_{0}^{u^{*}}\cP^{2}(u)du, \label{Z000}\\
Z_{1}&=-\frac{\log \cP'(u^{*})}{12}. \label{Z111}
\end{align}
We see that these formulae coincide with (\ref{Z0F},\ref{Z1F}).
In the next  sections we will use this formula for the torus
partition function in order to obtain expressions for correlation numbers.

\section{Evaluation of correlation numbers in genus one in KdV frame}

In the scaling limit near the $p$-critical point the partition
function of the One-matrix model on torus can be described in terms
of the solution of the ``string equation''
\begin{align}
\cP(u)=0, \label{Str0}
\end{align}
where $\cP(u)$ is the polynomial of degree $p+1$ ($p$ is natural
number)
\begin{align}
\cP(u)=u^{p+1}+t_{0}u^{p-1}+\sum\limits_{k=1}^{p-1}t_{k}u^{p-k-1},
\label{Pol}
\end{align}
with the parameters $t_{k}$ controling the deviation from the
$p$-critical point.The singular part of the partition function on
torus in the Matrix Models  $Z_{1}(t_{0}, t_{1},...t_{p-1})$ can be
described according to (\ref{Z111}), as
\begin{align}
Z_{1} =-\frac{\log \cP'(u^{*})}{12} , \label{StS}
\end{align}
where $u^{*}=u^{*}(t_{0},t_{1},...,t_{p-1})$ is the suitably chosen
root of the polynomial (\ref{Pol}). Also introduce $u_{c} =
u^{*}(t_{0},0,...,0)=\sqrt{-t_{0}}$. The correlation numbers are
expressed through the formula
\begin{align}
\langle O_{k_{1}}...O_{k_{n}} \rangle_{1} = \left.\frac{\partial^{n}
Z_{1}}{\partial t_{k_{1}}...\partial
t_{k_{n}}}\right|_{t_{1}=...=t_{p-1}=0},
\end{align}
where the index $\langle\;\rangle_{1}$ denotes the correlation
numbers on torus.  The first three correlation numbers are (see
Appendix D)
\begin{align}
&\langle O_{k}\rangle_{1} = \frac{p+k}{24}u_{c}^{-k-2}, \notag\\
&\langle O_{k_{1}}O_{k_{2}}\rangle_{1}=
\frac{(p+2+k_{1}+k_{2})(k_{1}+k_{2})+2p-k_{1}k_{2}}{48}u_{c}^{-k_{1}-k_{2}-4}, \notag\\
&\langle
O_{k_{1}}O_{k_{2}}O_{k_{3}}\rangle_{1}=\frac{1}{96}\left(\frac{2k^{3}}{3}+\frac{k_{i}^{3}}{3}+(p+4)k^2
+2k_{i}^{2}+(6p+8)k +8p
-2k_{1}k_{2}k_{3}\right)u_{c}^{-k_{1}-k_{2}-k_{3}-6},
\end{align}
where $k=k_{1}+k_{2}+k_{3}$,$\;\;
k_{i}^{2}=k_{1}^{2}+k_{2}^{2}+k_{3}^{2}$, and $
k_{i}^{3}=k_{1}^{3}+k_{2}^{3}+k_{3}^{3}$.

\section{Comparison with Topological Gravity }
In paper \cite{EW} E.Witten gave the definition of 2d Topological
gravity.  The recursion relation between correlation numbers  has
been derived in \cite{EW} by studying intersection theory. Using
this relation Witten  computed correlation numbers in genus zero and
checked their coincidence with expressions for correlation numbers
in  One-matrix model. This resemblance lead Witten to the conjecture
about the equivalence between Topological gravity and One-matrix
model, what was proved later by M.Kontsevich \cite{MK}.

In this section we show that the same recursion relation in
genus one as well as in genus zero holds also in One-matrix model.
Our apprach uses the explicit expression for the partion function of
One-matrix model and differs from that used by Kontsevich.

Let $\langle\; \rangle _{0}$ and $\langle\; \rangle _{1}$ denote the
genus zero and genus one correlation numbers. The recursion
relations between correlation numbers from \cite{EW} look as follows
\begin{align}
&\langle \sigma_{k_{1}}\sigma_{k_{2}}...\sigma_{k_{s}} \rangle_{0}
=k_{1} \sum_{S=X \cup Y} \langle \sigma_{k_{1}-1}\prod _{i \in X}
\sigma_{k_{i}} \sigma_{0} \rangle_{0} \langle \sigma_{0}\prod_{j \in
Y}\sigma_{k_{j}}\sigma_{k_{s-1}}\sigma_{k_{s}} \rangle_{0}, \label{r1}\\
&\langle \sigma_{k_{1}}\sigma_{k_{2}}...\sigma_{k_{s}} \rangle_{1} =
\frac{1}{12}k_{1}\langle
\sigma_{k_{1}-1}\sigma_{k_{2}}...\sigma_{k_{s}} \sigma_{0}\sigma_{0}
\rangle_{0} + k_{1} \sum_{S = X \cup Y} \langle
\sigma_{k_{1}-1}\prod_{i\in X} \sigma_{k_{i}}\sigma_{0}
\rangle_{0}\langle \sigma_{0}\prod_{j\in Y} \sigma_{k_{j}}
\rangle_{1},  \label{r2}
\end{align}
where $\sigma_{k} \leftrightarrow  O_{p-k-1}$ and the symbol
$\sum_{S=X \cup Y} $ represent a sum over all decomposition of
$S=\{2,3,...,s\}$ as a union of the two sets $X$ and $Y$.

Witten has shown that the relations (\ref{r1}) and (\ref{r2}) are
fulfilled for correlation numbers $\langle
\sigma_{k_{1}}...\sigma_{k_{s}} \rangle_{0,1}$  of the more general
theory which depends on relevant parameters $\{a_{k}\}$ ( $a_{k}
\leftrightarrow  t_{p-k-1}$) in such way that expectation values of
any observable $N$ obey
\begin{align}
&\frac{\partial}{\partial a_{k}}\langle N\rangle = \langle
\sigma_{k} N \rangle, \label{as3}
\end{align}
if the following relations
\begin{align}
&\langle \sigma_{k_{1}}\sigma_{k_{2}}\sigma_{k_{3}} \rangle_{0} =
k_{1}\langle \sigma_{k_{1}-1}\sigma_{0} \rangle_{0}\langle
\sigma_{0}\sigma_{k_{2}}\sigma_{k_{3}} \rangle_{0}\label{as1},\\
&\langle \sigma_{k} \rangle_{1} = \frac{1}{12}k \langle \sigma_{k-1}
\sigma_{0}\sigma_{0} \rangle_{0} + k\langle \sigma_{k-1} \sigma_{0}
\rangle_{0}\langle \sigma_{0} \rangle_{1}\label{as2}
\end{align}
hold.

In One-matrix model the correlation numbers with arbitrary
$\{t_{k}\}$ are given by the formula
\begin{align}
&\langle O_{k_{1}}...O_{k_{n}} \rangle_{0,1} = \frac{\partial^{n}
Z_{0,1}}{\partial t_{k_{1}}...\partial t_{k_{n}}}.
\end{align}
Therefore the assumption (\ref{as3}) is holds automatically. The
fulfillment of (\ref{as1}) for correlation numbers in genus
zero was checked by A.B. Zamolodchikov \cite{Z}.

Below we check (\ref{as1}) and (\ref{as2}) using explicit
expressions for $Z_{0}$ and $Z_{1}$ from (\ref{Z000}) and
(\ref{Z111}). In terms of observables $O_{k}$ in One-matrix model,
the expressions (\ref{as1}) and (\ref{as2}) writes ($\sigma_{k}
\leftrightarrow O_{p-k-1}$)
\begin{align}
&\langle O_{p-k_{1}-1}O_{p-k_{2}-1}O_{p-k_{3}-1} \rangle_{0} =
k_{1}\langle O_{p-k_{1}}O_{p-1} \rangle_{0}\langle
O_{p-1}O_{p-k_{2}-1}O_{p-k_{3}-1} \rangle_{0}\label{asO1},\\
&\langle O_{p-k-1} \rangle_{1} = \frac{1}{12}k \langle O_{p-k}
O_{p-1}O_{p-1} \rangle_{0} + k\langle O_{p-k} O_{p-1}
\rangle_{0}\langle O_{p-1} \rangle_{1}.  \label{asO2}
\end{align}
From formula (\ref{Z000}) at arbitrary $\{t_{k}\}$ one can get
\begin{align}
\langle O_{k_{1}}O_{k_{2}}\rangle_{0} =
\frac{\partial^{2}Z_{0}}{\partial t_{k_{1}}\partial t_{k_{2}}} =
\frac{(u^{*})^{2p-k_{1}-k_{2}-1}}{2p-k_{1}-k_{2}-1}, \quad \langle
O_{k_{1}}O_{k_{2}}O_{k_{3}}\rangle_{0}
=\frac{\partial^{3}Z_{0}}{\partial t_{k_{1}}\partial
t_{k_{2}}\partial t_{k_{3}}}=
-\frac{(u^{*})^{3p-k_{1}-k_{2}-k_{3}-3}}{\cP'(u^{*})}. \label{CorO}
\end{align}
Thus it is easy to see from (\ref{CorO}) that equality (\ref{asO1})
is fulfilled.  From the torus partition function (\ref{Z111}) also
at arbitrary $\{t_{k}\}$ we get
\begin{align}
\langle  O_{k} \rangle_{1} =\frac{\partial Z_{1}}{\partial
t_{k}}=-\frac{p-k-1}{12\cP'(u^{*})}(u^{*})^{p-k-2}
+\frac{\cP''(u^{*})}{12(\cP'(u^{*}))^{2}}(u^{*})^{p-k-1}.
\label{O11}
\end{align}
The expressions (\ref{CorO}) and (\ref{O11}) indeed satisfy  the
equation (\ref{asO2}). Consequently we have proved that the
correlation numbers in One-matrix model in KdV frame satisfy the
recurrence relation (\ref{r1}) and (\ref{r2}), assuming replacement
$\sigma_{k}\to O_{p-k-1}$.

\section{Evaluation of correlation numbers in the CFT frame}
The CFT frame is defined by a different set of  parameters
$\{\lambda_{k}\}$, which are associated with $\{t_{k}\}$ by
"resonance" transformation \cite{BZ2}. As it was shown in \cite{BZ2}
after "resonance" transformation the polynomial $\cP(u,\{t_{k}\})$
from (\ref{Pol}) up to  the  factor
$\frac{(p+1)!}{(2p-1)!!}u_{c}^{p+1}$ takes the form
\begin{align}
Q(x,\{\lambda_{k}\})
=\sum\limits_{n=0}^{\infty}\sum\limits_{k_{1}...k_{n}=1}^{p-1}
\frac{\lambda_{k_{1}}...\lambda_{k_{n}}}{n!}\frac{d^{n-1}}
{dx^{n-1}}L_{p-\sum k_{i}-n}(x),
\end{align}
where $x=u/u_{c}$, $u_{c}$ is $u^{*}$ at
$\lambda_{1},...,\lambda_{p-1}=0$ and $L_{n}(x)$ are the  Legendre
polynomials. We also assume that
$\left(\frac{d}{dx}\right)^{-1}L_{p} = \int L_{p} dx
=\frac{L_{p+1}-L_{p-1}}{2p+1}$. Below we use the notation
\begin{align}
Q_{k_{1}...k_{n}}(x)=\frac{d^{n-1}} {dx^{n-1}}L_{p-\sum k_{i}-n}(x),
\quad Q_{0}(x)= \frac{L_{p+1}-L_{p-1}}{2p+1}.
\end{align}
The correlation numbers are expressed through the formula
\begin{align}
\langle \mathcal{O}_{k_{1}}...\mathcal{O}_{k_{n}} \rangle_{1} =
\left.\frac{\partial^{n} Z_{1}}{\partial \lambda_{k_{1}}...\partial
\lambda_{k_{n}}}\right|_{\lambda_{1}=...=\lambda_{p-1}=0}.
\label{corCFT}
\end{align}
If we calculated correlation numbers for the partition function in
genus zero, inserting the polynomial $Q(x,\{\lambda_{k}\})$ instead
of $P(x,{t_{k}})$ in (\ref{Z000}) and (\ref{Z111}) we would
\textsc{}obtained results in \cite{BZ2}.

Let us compute the partition function in genus one. Thus taking into
account the common formulas for correlation numbers (\ref{O}) and
(\ref{OO}) in Appendix D and using some values for Legendre
polynomials and consequently for polynomial $Q(x,\{\lambda_{k}\})$
in critical point ($x=1$)
\begin{align}
&Q'(1)=1,\quad Q''(1) =\frac{p(p+1)}{2},\quad Q'''(1) =
\frac{(p+2)(p+1)p(p-1)}{8}, \quad Q_{k_{i}}=1,\notag\\
&Q'_{k_{i}}(1) =\frac{(p-k_{i})(p-k_{i}-1)}{2},\quad
Q''_{k_{i}}(1)=\frac{1}{8}\prod_{r=1}^{4}(p-r-k_{i}+2),\notag\\
&Q'_{k_{i}k_{j}}(1) =\frac{1}{8}\prod_{r=1}^{4}(p-r-k_{i}-k_{j}+1),
\end{align}
one can obtain  from (\ref{corCFT}) the first two correlation
numbers of the partition function in genus one in CFT frame
\begin{align}
&\langle \mathcal O_{k} \rangle_{1}
= \frac{(2p-k)(k+1)}{24}, \notag\\
&\langle \mathcal O_{k_{1}}\mathcal O_{k_{2}} \rangle_{1}
=-\frac{1}{24}(1+k_{1})(1+k_{2})((k_{1}+k_{2}-2p+2)(k_{1}+k_{2})-k_{1}k_{2}-4p).
\end{align}

\section{Conclusion}

In this paper we have derived the torus partition function $Z_{1}$
in  $p$-critical One-matrix model.  Using the explicit expression
for the partition function in genus one we compute the correlation
numbers in KdV, as well as in CFT frames.

We show the fulfillment of  recurrence relation for correlation
numbers in OMM in KdV frame in genus one, which are the same as that
in 2d Topological gravity .

The results in CFT frame should be compared against the correlation
numbers in the Minimal Liouville gravity, which have not been
computed yet. We expect the  coincidence in genus one similarly that
was observed on sphere \cite{MSS, BZ2, BAlZ} and on disk \cite{BR}.

\section{Acknowledgements}
We are grateful to V. Belavin, M. Bershtein, M. Lashkevich, Ya.
Pugai and A. Zamolodchikov for useful discussions.

This work was supported  by Federal Program Scientific-Pedagogical
Personnel of Innovation Russia (contract No. 02.740.11.5165) and by
grant Scientific Schools 6501.2010.2.
A.B. was supported also by RFBR initiative interdisciplinary
project grant 09-02-12446-ofi-m and RBRF-CNRS grant PICS-09-02-91064.

\section*{Appendix}
\appendix

\section{}
For our aim in section 3 we use Euler-Maclorein formula \cite{FG}.
It helps express summation of discrete function through integration
of this function and some other terms.

Let function $f(x)$ is considered in section $[a,b]$. Let
$h=\frac{b-a}{n}$, where $n$  is natural number, then
\begin{equation}
\sum_{k=1}^{n}f(a +(k-1) h) = \frac{1}{h}
\int_{a}^{b}f(x)dx-\frac{1}{2}(f(b)-f(a))+
\sum_{m=1}^{\infty}h^{2m-1}\frac{B_{2m}}{(2m)!}(f^{(2m-1)}(b)-f^{(2m-1)}(a)),
\end{equation}
where $B_{m}$ are Bernoulli numbers ($B_{0}=1, B_{1}=-\frac{1}{2},
B_{2}= \frac{1}{6}$). In our analysis of torus partition function we
only use terms in Euler-Maclorein formula up to $m=1$.

\section{}
In this section we deal with part of the sum from (\ref{TW})
\begin{align}
\tilde{W}_{n} = 2nv_{n}\sum_{\{\sigma_{2n-1}\}}
R_{k+m_{1}}\cdot...\cdot R_{k+m_{n}}, \label{sum1}
\end{align}
where $\{\sigma_{2n-1}\}$ denotes all paths\, of $2n-1$ steps ($n-1$
steps up and $n$ steps down) starting at $k$ and ending at $k-1$.
Each step down from $m$ to $m-1$ receives a factor of $R_{m}$ and
each step up receives a factor of unity.

We assume the existence of smooth function $R(\xi,N)$ of variable $\xi
\in [0,1]$, such that $R(\frac{k}{N},N)=R_{k}$. Thus $R(\xi+m/N,N)$
and $\tilde{W}_{n}$ have the Taylor expansion
\begin{align}
&R(\xi+m/N,N) =
R(\xi,N)+\frac{m}{N}R_{\xi}(\xi,N)+\frac{m^{2}}{2N^{2}}R_{\xi\xi}(\xi,N)+O\big(\frac{1}{N^{3}}\big),\\
&\tilde{W}_{n} = W_{n}
+\frac{1}{N}W_{1n}+\frac{1}{N^{2}}W_{2n}+O\big(\frac{1}{N^{3}}\big).
\label{WA1}
\end{align}
Since
\begin{multline}
R(\xi+\frac{m_{1}}{N},N)\cdot... \cdot R(\xi+\frac{m_{n}}{N},N)
=\\
=R^{n}+ \frac{R^{n-1}R_{\xi}}{N}\sum_{i=1}^{n}m_{i}+ \frac{
R^{n-1}R_{\xi\xi}}{2N^{2}}\sum_{i=1}^{n}m^{2}_{i}+\frac{R^{n-2}R_{\xi}^{2}}{N^{2}}
\sum_{i<j}^{n}m_{i}m_{j}.
\end{multline}
Thus from (\ref{WA1}) we obtain
\begin{align}
&W_{n}= 2nv_{n}R^{n}\sum_{\{\sigma_{2n-1}\}}1,\quad
W_{1n}=2nv_{n}R^{n-1}R_{\xi}\sum_{\{\sigma_{2n-1}\}}\sum_{i=1}^{n}m_{i},
\notag \\
&W_{2n} = 2nv_{n}R^{n-2}R_{\xi}^{2}
\sum_{\{\sigma_{2n-1}\}}\sum_{i<j}^{n}m_{i}m_{j}+nv_{n}R^{n-1}R_{\xi\xi}
\sum_{\{\sigma_{2n-1}\}}\sum_{i=1}^{n}m^{2}_{i}.
\end{align}
Therefore we need to evaluate the sums
\begin{align}
P_{n}=\sum_{\{\sigma_{2n-1}\}}1,\quad M_{n}=
\sum_{\{\sigma_{2n-1}\}}\sum_{i=1}^{n}m_{i}, \quad
A_{n}=\sum_{\{\sigma_{2n-1}\}}\sum_{i=1}^{n}m^{2}_{i},\quad
B_{n}=\sum_{\{\sigma_{2n-1}\}}\sum_{i<j}^{n}m_{i}m_{j}.
\end{align}

First of all, it is easy to see that $P_n$, which is the total
number of paths, is given by the binomial coefficient~$C^n_{2n-1}$.

Thus we have for $P_{n}=C_{2n-1}^{n}$, and one can writes
\begin{align}
&P_{n}=\sum_{\{\sigma_{2n-1}\}}1=C_{2n-1}^{n},\notag\\
&M_{n}=\sum_{\{\sigma_{2n-1}\}}\sum_{i=1}^{n}m_{i}=
\sum_{x,y=0}^{n-1}(y-x)C_{x+y}^{x}C_{2n-2-x-y}^{n-1-x}=0 , \notag\\
&A_{n}=\sum_{\{\sigma_{2n-1}\}}\sum_{i=1}^{n}m^{2}_{i}=
\sum_{x,y=0}^{n-1}(y-x)^{2}C_{x+y}^{x}C_{2n-2-x-y}^{n-1-x}= \frac{n(n-1)}{3}C_{2n-1}^{n} ,\notag\\
&B_{n}=\sum_{\{\sigma_{2n-1}\}}\sum_{i<j}^{n}m_{i}m_{j}=
\sum_{x,y=0}^{n-1}(y-x)C_{2n-2-x-y}^{n-1-x}\sum_{x_{1}=0}^{x-1}
\sum_{y_{1}=0}^{y}(y_{1}-x_{1})C_{x_{1}+y_{1}}^{x_{1}}C_{x-1+y-y_{1}-x_{1}}^{y-y_{1}}=\notag\\
&\qquad \qquad \qquad \qquad \qquad\qquad \qquad \qquad \qquad
\qquad \qquad \qquad =\frac{n(n-1)(n-2)}{12}C_{2n-1}^{n}.
\end{align}
Thus we obtain for $W = \sum_{n=0}^{p+1}2nv_{n}W_{n}, \; W_{1}
=\sum_{n=0}^{p+1}2nv_{n}W_{1n} $ and  for $W_{2} =
\sum_{n=0}^{p+1}2nv_{n}W_{2n}$, following expressions
\begin{align}
&W(R(\xi,N))=\sum_{n=1}^{p+1}\frac{(2n)!}{n!(n-1)!}v_{n}R^{n}(\xi,N),\notag\\
&W_{1}(R(\xi,N))=0,\notag\\
&W_{2}(R(\xi,N))=\frac{RR_{\xi\xi}}{6}W''(R(\xi,N))
+\frac{RR_{\xi}^{2}}{12}W'''(R(\xi,N)).
\end{align}

\section{}
In this section we  solve the recurrence relation for
$S_{k}[u,u',u'',...]$
\begin{align}
\frac{dS_{k+1}}{dx} =  u\frac{dS_{k}}{dx} +
\frac{1}{2}u_{x}S_{k}+\frac{\varepsilon^{2}}{4}\frac{d^{3}S_{k}}{dx^{3}},
\label{Arr}
\end{align}
with the boundary conditions $S_{0}=\frac{1}{2}$ and $S_{k}$($k\neq
0$) vanish at $u=0$. From the form of equation (\ref{Arr}) it
follows that $S_{k}$ is expanded into series of $\varepsilon$:
\begin{align}
S_{k} = S_{k}^{0}+\varepsilon
S_{k}^{1}+...+\varepsilon^{l}S_{k}^{l}+....
\end{align}
and $l$-th term in this expansion contains the common number of
derivatives equal to $l$. In order to obtain the partition  function
of genus one we can limit this expansion up to the first three
terms, thus we have
\begin{align}
S_{k} = P_{k}(u)+\varepsilon
M_{k}(u)u_{x}+\varepsilon^{2}(A_{k}(u)u_{xx}+B_{k}(u)u_{x}^{2})+O(\varepsilon^{3}).
\end{align}
For the r.h.s. of the recurrence relation (\ref{Arr}) we obtain
\begin{align}
&u\frac{dS_{k}}{dx}= P'_{k} uu_{x} +\varepsilon \left( M'_{k}
uu_{x}^{2}+M_{k}uu_{xx}\right)+
\varepsilon^{2}\left(A_{k}uu_{xxx}+B'_{k}
uu_{x}^{3}+\left(2B_{k}+A'_{k}\right)uu_{x}u_{xx}\right)+O(\varepsilon^{3}),\notag\\
&\frac{1}{2}u_{x}S_{k}=\frac{1}{2}P_{k}u_{x}+\varepsilon
\frac{M_{k}}{2}u_{x}^{2}+\varepsilon^{2}\left(\frac{A_{k}}{2}u_{x}u_{xx}+\frac{B_{k}}{2}u_{x}^{3}\right)
+O(\varepsilon^{3}),\notag\\
&\frac{\varepsilon^{2}}{4}\frac{d^{3}S_{k}}{dx^{3}}=\varepsilon^{2}
\left(\frac{1}{4}P'''_{k}u_{x}^{3}+\frac{3}{4}P''_{k}u_{x}u_{xx}
+\frac{1}{4}P'_{k}u_{xxx}\right) + O(\varepsilon^{3}),
\end{align}
and for the l.h.s
\begin{multline}
\frac{dS_{k+1}}{dx} = P'_{k+1}u_{x} +\varepsilon\left(
M'_{k+1}u_{x}^{2}+M_{k+1}u_{xx}\right)+\\
+\varepsilon^{2}\left(A_{k+1}u_{xxx}+ B'_{k+1}
u_{x}^{3}+\left(2B_{k+1}+A'_{k+1}\right)u_{x}u_{xx}\right)+O(\varepsilon^{3}).
\end{multline}
In the order $\varepsilon^{0}$ we have one equation
\begin{align}
 P'_{k+1} =
P'_{k}u +\frac{1}{2}P_{k},
\end{align}
with boundary condition $P_{0}=\frac{1}{2}$. And if we make
substitution $P_{k} =p_{k}u^{k}$, where $p_{k}$ is constant, we
obtain recurrence relation for $p_{k}$
\begin{align}
p_{k+1} =\frac{2k+1}{2(k+1)}p_{k}.
\end{align}
Solving this equation we find that $p_{k}=
\frac{C_{2k}^{k}}{2^{2k+1}} $, where $C_{2k}^{k} =
\frac{(2k)!}{k!k!}$ is the binomial coefficient. Thus we derive that
$P_{k}(u) =\left(\frac{C_{2k}^{k}}{2^{2k+1}}\right) u^{k} $.

For order $\varepsilon$ we have two equations
\begin{align}
\begin{cases}
 M'_{k+1}=  M'_{k}u+\frac{1}{2}M_{k}\\
M_{k+1}= M_{k}u
\end{cases},\label{M}
\end{align}
with boundary condition $M_{0}=0$. The system of equations (\ref{M})
has only one solution $M_{k}=0$. It easy to see that already from
$M_{k}=0$ it follows, that all terms of odd order in $\varepsilon$
will be qual to zero.

In order $\varepsilon^{2}$ we have system of three equations
\begin{align}
\begin{cases}
A_{k+1}= A_{k}u+\frac{1}{4} P'_{k},\\
 B'_{k+1}=
B'_{k} u+\frac{1}{2}B_{k}+\frac{1}{4}
P'''_{k},\\
2B_{k+1}+A'_{k+1} = \left(2B_{k}+ A'_{k}\right)u +\frac{1}{2}A_{k}
+\frac{3}{4}P''_{k},
\end{cases}\label{B2}
\end{align}
with boundary conditions $A_{0}=B_{0}=0$. Let us solve the first
one.
\begin{align}
A_{k+1}=A_{k}u+\frac{1}{4}P'_{k}. \label{A}
\end{align}
In order to get rid of the heterogeneous component
$\frac{1}{4}\frac{\partial P_{k}}{\partial u}$, we will look for
$A_{k}$ in form $A_{k}
=a_{k}\frac{k(k-1)}{4}\left(\frac{C_{2k}^{k}}{2^{2k+1}}\right)u^{k-2}$,
where $a_{k}$ are constants, and we obtain for $a_{k}$ from
(\ref{A})
\begin{align}
a_{k+1}(k+1/2)-a_{k}(k-1)=1.
\end{align}
Solution for this equation is $a_{k} =  \frac{2}{3}$, thus we have
$A_{k} =
\frac{k(k-1)}{6}\left(\frac{C_{2k}^{k}}{2^{2k+1}}\right)u^{k-2}$.
For the second equation from (\ref{B2}) we put $B_{k}
=b_{k}\frac{k(k-1)(k-2)}{4}\left(\frac{C_{2k}^{k}}{2^{2k+1}}\right)u^{k-3}$,
then  we derive equation for the constants $b_{k}$:
\begin{align}
b_{k+1}(k+1/2)-b_{k}(k-5/2)=1,
\end{align}
and obtain $b_{k}= \frac{1}{3}$, therefore $B_{k}
=\frac{k(k-1)(k-2)}{12}\left(\frac{C_{2k}^{k}}{2^{2k+1}}\right)
u^{k-3}$. We can see that these solutions for $A_{k}$ and $B_{k}$
satisfy the third equation in (\ref{B2}).

Summarizing all results, we have
\begin{align}
&P_{k}= \left(\frac{C_{2k}^{k}}{2^{2k+1}}\right)u^{k},\quad M_{k}=0,  \notag\\
&A_{k}=\frac{k(k-1)}{6}\left(\frac{C_{2k}^{k}}{2^{2k+1}}\right)
u^{k-2} = \frac{1}{6}P''_{k}\notag\\
&B_{k}=\frac{k(k-1)(k-2)}{12}\left(\frac{C_{2k}^{k}}{2^{2k+1}}\right)
u^{k-3}=\frac{1}{12}P'''_{k},
\end{align}
therefore one can writes $S_{k}$ as follows
\begin{align}
S_{k} = P_{k} +\varepsilon^{2}\left(\frac{1}{6}P''_{k}u_{xx}
+\frac{1}{12}P'''_{k}u_{x}^{2}\right) +O(\varepsilon^{4}).
\end{align}

\section{ }

The singular part of the partition function on torus $Z_{1}(t_{0},
t_{1},...t_{p-1})$ is
\begin{align}
Z_{1} =-\frac{\log \cP'(u^{*})}{12},
\end{align}
where $u^{*}=u^{*}(t_{0},t_{1},...,t_{p-1})$ is the suitably chosen
root of the polynomial
\begin{align}
\cP(u)=u^{p+1}+t_{0}u^{p-1}+\sum\limits_{k=1}^{p-1}t_{k}u^{p-k-1}.
\label{PolA}
\end{align}
The correlation numbers are expressed through the formula
\begin{align}
\langle O_{k_{1}}...O_{k_{n}} \rangle_{1} = \left.\frac{\partial^{n}
Z_{1}}{\partial t_{k_{1}}...\partial
t_{k_{n}}}\right|_{t_{1}=...=t_{p-1}=0}.
\end{align}
Thus in common form first two correlation numbers are (denote
$\cP_{k} =\partial \cP /\partial t_{k}$)
\begin{align}
&\langle O_{k_{1}} \rangle_{1} =
-\frac{1}{12}\left(\frac{\cP'_{k_{1}}}{\cP'}
-\frac{\cP''\cP_{k_{1}}}{(\cP')^{2}}\right),\label{O}\\
&\langle O_{k_{1}}O_{k_{2}} \rangle_{1}
=-\frac{1}{12}\left(\frac{\cP'_{k_{1}k_{2}}}{\cP'}-\frac{\cP''_{k_{1}}\cP_{k_{2}}+\cP''_{k_{2}}\cP_{k_{1}}+\cP'_{k_{1}}
\cP'_{k_{2}}+\cP''\cP_{k_{1}k_{2}}}{(\cP')^{2}}+\right.\notag\\
&\left.\qquad\qquad\qquad+\frac{2\cP''\cP'_{k_{1}}\cP_{k_{2}}+2\cP''\cP'_{k_{2}}
\cP_{k_{1}}+\cP'''\cP_{k_{1}}\cP_{k_{2}}}{(\cP')^{3}}-
\frac{2(\cP'')^{2}\cP_{k_{1}}\cP_{k_{2}}}{(\cP')^{4}}\right).
\label{OO}
\end{align}
And the third correlation number is (denote
$\cP_{i}=\cP_{k_{i}}=\partial \cP /\partial k_{i}$)
\begin{align}
\langle O_{k_{1}}O_{k_{2}}O_{k_{3}}
\rangle_{1}&=-\frac{1}{12}\left(\frac{\cP'_{123}}{\cP'}-\frac{\cP''_{(12}\cP_{3)}+\cP'_{(12}\cP_{3)}+
\cP''_{(1}\cP_{23)}+\cP''\cP_{123}}{(\cP')^{2}}\right.+\notag\\
&+\frac{1}{(\cP')^{3}}\left(2\cP''\cP'_{(12}\cP_{3)}+\cP'''_{(1}\cP_{2}\cP_{3)}+
2\cP''_{(1}\cP'_{2}\cP_{3)}+\cP'''\cP_{(12}\cP_{3)}+2\cP''\cP'_{(1}\cP_{23)}+2\cP'_{1}\cP'_{2}\cP'_{3}\right)-\notag\\\
&-\frac{1}{(\cP')^{4}}\left(4\cP''\cP''_{(1}\cP_{2}\cP_{3)}+6\cP''\cP'_{(1}\cP'_{2}\cP_{3)}+
2(\cP'')^{2}\cP_{(12}\cP_{3)}+3\cP'''\cP'_{(1}\cP_{2}\cP_{3)}+\cP''''\cP_{1}\cP_{2}\cP_{3}\right)+\notag\\
&\left.+\frac{1}{(\cP')^{5}}\left(8(\cP'')^{2}\cP'_{(1}\cP_{2}\cP_{3)}+7\cP'''\cP''\cP_{1}\cP_{2}\cP_{3}\right)
-\frac{8(\cP'')^{3}\cP_{1}\cP_{2}\cP_{3}}{(\cP')^{6}}\right),
\label{OOO}
\end{align}
where parentheses denote symmetrization (for instance
$\cP''_{(12}\cP_{3)}
=\cP''_{12}\cP_{3}+\cP''_{23}\cP_{1}+\cP''_{31}\cP_{2}$).

In KdV critical point i.e. $t_{1}=...=t_{p-1}=0$, we have
$u_{c}=u_{*}(t_{0},0,...,0)=\sqrt{-t_{0}}$, and for differen
derivatives of polynomial $\cP(u)$ from (\ref{PolA}) one can get
\begin{align}
&\cP'(u_{c}) =2u_{c}^{p}, \quad \cP''(u_{c}) =
2(2p-1)u_{c}^{p-1},\quad
\cP'''(u_{c}) = 6(p-1)^{2}u_{c}^{p-2}\notag\\
&\cP_{k_{i}}(u_{c}) = u_{c}^{p-k_{i}-1},\quad
\cP'_{k_{i}}(u_{c})=(p-k_{i}-1)u_{c}^{p-k_{i}-2},
\;\;\cP''_{k_{i}}(u_{c})=(p-k_{i}-1)(p-k_{i}-2)u_{c}^{p-k_{i}-3},\notag\\
&\cP_{k_{i}k_{j}}(u_{c})=0.   \label{ef}
\end{align}
Thus after substitution the expressions (\ref{ef}) in formulas
(\ref{O}), (\ref{OO}) and (\ref{OOO}) we obtain
\begin{align}
&\langle O_{k}\rangle_{1} = \frac{p+k}{24}u_{c}^{-k-2}, \notag\\
&\langle O_{k_{1}}O_{k_{2}}\rangle_{1}=
\frac{(p+2+k_{1}+k_{2})(k_{1}+k_{2})+2p-k_{1}k_{2}}{48}u_{c}^{-k_{1}-k_{2}-4}, \notag\\
&\langle
O_{k_{1}}O_{k_{2}}O_{k_{3}}\rangle_{1}=\frac{1}{96}\left(\frac{2k^{3}}{3}+\frac{k_{i}^{3}}{3}+(p+4)k^2
+2k_{i}^{2}+(6p+8)k +8p
-2k_{1}k_{2}k_{3}\right)u_{c}^{-k_{1}-k_{2}-k_{3}-6},
\end{align}
where $k=k_{1}+k_{2}+k_{3}$,$\;\;
k_{i}^{2}=k_{1}^{2}+k_{2}^{2}+k_{3}^{2}$, and $
k_{i}^{3}=k_{1}^{3}+k_{2}^{3}+k_{3}^{3}$.

\end{document}